\documentclass[prd,a4paper,twocolumn]{revtex4}

\usepackage{graphicx}

\begin{document}

\title{Observed Properties of Dark Matter: dynamical studies of dSph galaxies}

\author{G Gilmore$^1$, M Wilkinson$^1$, J Kleyna$^2$, A Koch$^4$,  Wyn Evans$^1$,
R.F.G. Wyse$^3$, E.K. Grebel$^4$}
\affiliation{1: Institute of Astronomy, Cambridge, UK; 2: Institute
for Astronomy, Honolulu, Hawaii, USA; 3: Johns Hopkins University, Baltimore,
USA; 4: Astronomical Institute, University of Basel, Switzerland}

\begin{abstract}

The Milky Way satellite dwarf spheroidal (dSph) galaxies are the
smallest dark matter dominated systems in the universe.  We have
underway dynamical studies of the dSph to quantify the shortest scale
lengths on which Dark Matter is distributed, the range of Dark Matter central
densities, and the density profile(s) of DM on small scales.
Current results suggest some surprises: the central DM
density profile is typically cored, not cusped, with scale sizes never
less than a few hundred pc; the central densities
are typically $10-20$GeV/cc; no galaxy is found with a dark mass halo less
massive than $\sim5.10^7M_{\odot}$. We are discovering many more dSphs, which
we are analysing to test the generality of these results.

\end{abstract}

\maketitle

\section{Introduction and Methodology}

Determining the nature of the dark matter is one of the key goals of
contemporary astronomy.  The extremely large mass to light ratios for
certain Local Group dwarf spheroidal (dSph) galaxies suggest that
these are the most dark matter dominated stellar systems known in the
Universe. Given the apparent absence of dark matter in globular
clusters (length scale $\sim10$pc), and the direct evidence that there
is no dark matter associated with the galactic thin disk (length scale
$\sim 100$pc; \cite{KG89,KG91})  the dSphs (characteristic
radii $\sim300+$ pc) are both the smallest systems to contain
dynamically significant quantities of dark matter, and the most nearby
systems where we may look for characteristic properties of the
distribution of dark matter (maximum density, etc ...), which may
provide some knowledge of its nature.  Additionally, simulations of
dark halo formation in a cosmological context predict that dSph
galaxies are (unmodified?) survivors from the earliest structures
formed, so that understanding their nature and structure also has
general implications for galaxy formation and evolution.

Is it possible reliably to determine the dark matter distribution in a
dSph galaxy? Fortunately, some dSphs are astrophysically simple, are
old, are apparently in equilibrium, have
no dynamically-significant gas, but contain a (small) number of stars,
which provide ideal kinematic collisionless tracer particles.  With
the new generation of wide-field multi-object spectrographs on 4-8m
telescopes, it is now viable to obtain sufficient high-quality
kinematic data to determine the gravitational potentials in all the
Galactic satellites.  There are two types of analysis feasible with
current datasets: a straighforward pressure-gravity balance analysis,
based on the velocity dispersion moment, using the Jeans' equations,
and an analysis of the full projected kinematic distribution function,
using the full information set but requiring more sophisticated
analyses. These analyses are made more robust, and simplified, by
stellar population studies of the dSph galaxies. Ultra-deep HST
imaging for Draco \cite{Gr} and UMi \cite{Wy} show
their stellar populations are indeed indistinguishable from those of
old metal-poor globular star clusters, and hence have  $M/L_V\sim2$.

The Jeans' equations for a spherical stellar system lead directly to a
model-independent mass estimator (see \cite{BT}
Eq. 4-55 \& 4-56) requiring only knowledge of the velocity anisotropy
and the true radial velocity dispersion $\langle v_r^2
\rangle$. Assuming spherical symmetry, it is straightforward to obtain
$\langle v_r^2 \rangle$ from the line of sight velocity dispersion
$\langle v^2 \rangle$ using Abel integrals. In applying the Jeans'
equation analyses, the spatially-binned dispersion profile is fit by
an appropriate smooth function (figure 1), while the light is fit by a
Plummer law, as this is an excellent fit to available star count data
in all available cases (cf \cite{K01}).

\begin{figure}[!h]
\begin{center}
\includegraphics[width=0.42\textwidth]{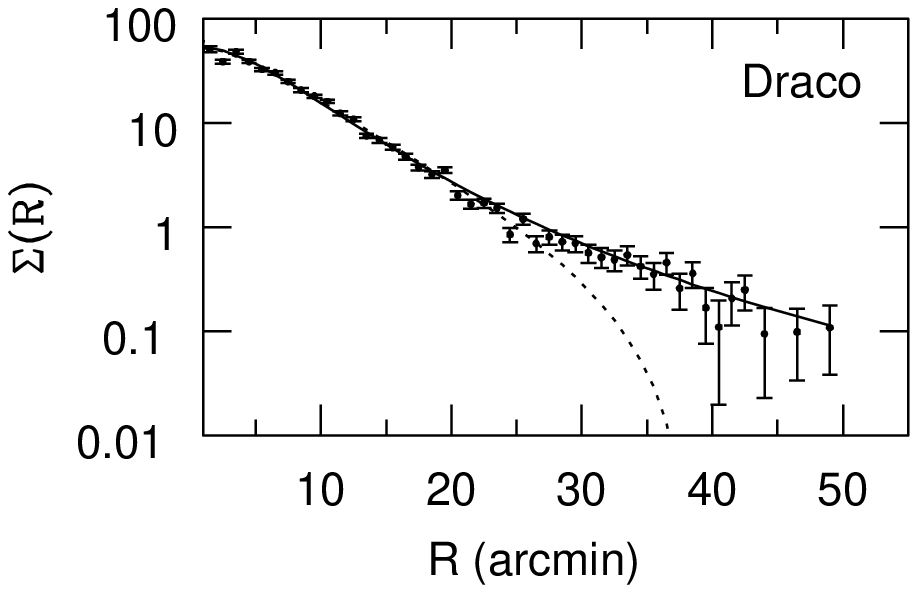}
\includegraphics[width=0.42\textwidth]{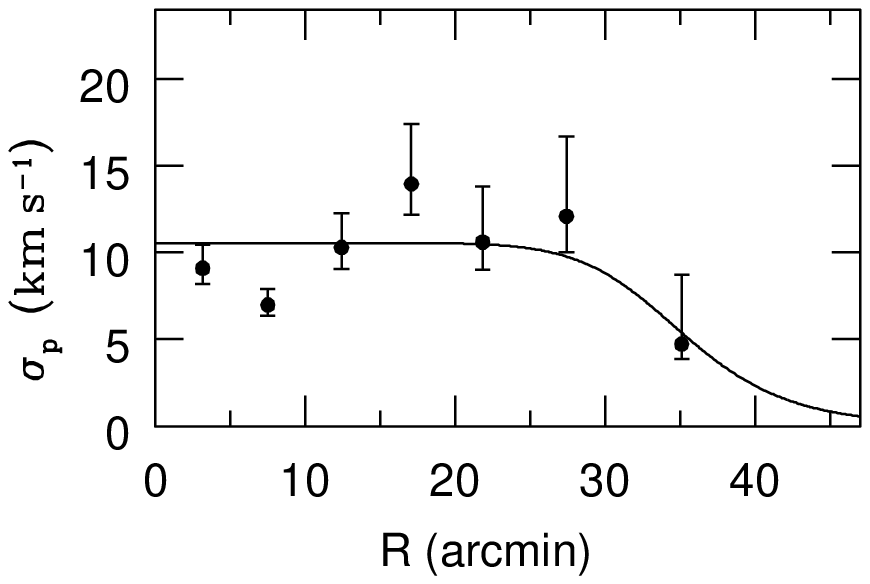}
\caption{Draco surface brightness profile, with a fitted Plummer
model, and the observed line of sight velocity dispersion
profile, from \cite{K01}.}
\end{center}
\end{figure}

The Jeans' equation moment analysis involves two radial functions, the
mass density and a possibly variable anisotropy in the stellar orbits,
this second generating an anisotropic stress tensor. It is this
anisotropic stress which gives the dSph galaxies their shapes, as
direct observational limits exclude any angular momentum support
against gravity. Equivalently, radially variable anisotropy is
degenerate with mass (figure 2) making any deductions on the inner
mass profile being cored or cusped in general model-dependent. Further
information is needed to break this degeneracy, and fortunately is
sometimes available, as we discuss below. In general however, full
multi-component distribution-function models are required to use the
information in the data to break this degeneracy.

\begin{figure}[!h]
\begin{center}
\includegraphics[width=0.42\textwidth]{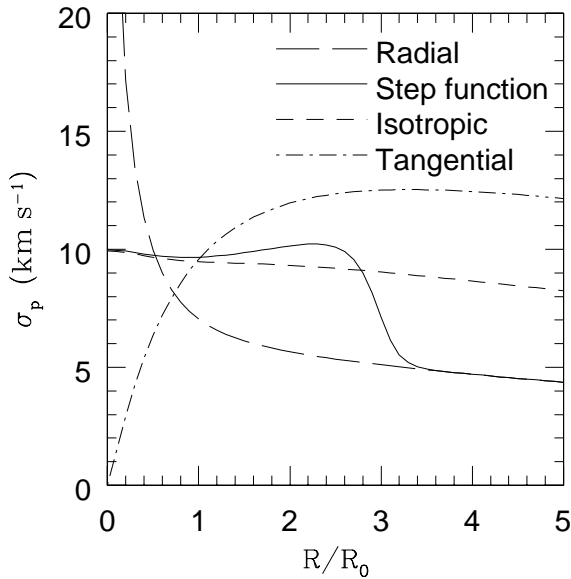}
\caption{An illustration of the possible effects of stellar orbital
anistropy variations on kinematic modelling, from \cite{W04}.}
\end{center}
\end{figure}

\section{Results of dynamical analyses}

Are dSph haloes cusped or cored? Debate continues to rage about
whether the cusped haloes always created in CDM simulations are in
conflict with observations of rotation curves of Low Suface Brightness
galaxies. The gas-free nature of dSphs makes them kinematically clean
systems in which to test theoretical predictions.  Stellar velocities
may also be used to place constraints on the steepness of any possible
central cusp, whether due to a black hole, the intrinsic physical
properties of the CDM (eg \cite{T02}), or possibly
even CDM as modified by a central black hole (eg \cite{rg}).

As a general result, in all cases with sufficient data we rule out
(King model) mass-follows-light models. King models are not an
adequate description of these galaxies, all of which have high
mass-to-light ratios (quoted in solar visual band units), and extended
dark matter halos.  Even in the inner regions mass does not follow
light, while including outer data commonly we find a most likely
global mass to light ratio which is very high, being for Draco $\sim
440$, 200 times greater than that for stars with a normal mass function
(Fig.~1).  The Draco halo models favoured by the data contain
significant amounts of mass at large radii, leading to the observed
flat to rising velocity dispersion profiles at intermediate to large
radii.

\begin{figure}[!h]
\begin{center}
\includegraphics[width=0.42\textwidth]{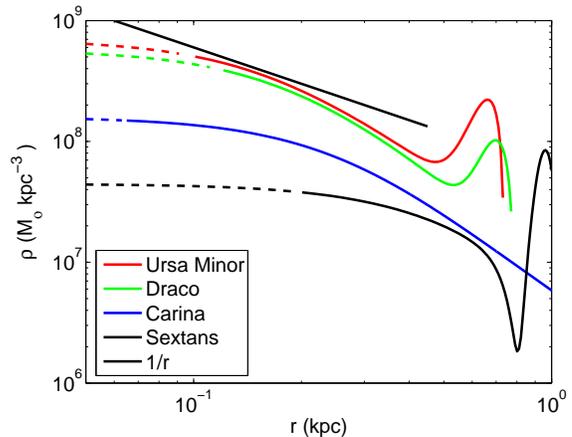}
\caption{Derived inner mass distributions from Jeans' eqn analyses for
four dSph galaxies. Also shown is a predicted $r^{-1}$ density
profile. The modelling is reliable in each case out to radii of log
(r)kpc$\sim0.5$. The unphysical behaviour at larger radii is explained
in the text. The general similarity of the four inner mass profiles is
striking.}
\end{center}
\end{figure}

In some cases (eg Draco, cf Figure 1) a decrease in the disersion is
apparent in the outermost data. This remains to be fully
understood. The decrease is apparently steeper than Keplerian, so does
not indicate the (first) detection of the outer limits of the
galaxy. It cannot indicate tidal perturbations by the Milky Way
galaxy, and in fact robustly excludes such complications, as tides
heat and do not cool (eg \cite{R06}). It may indicate a combination of
a radially-changing orbital anisotropy and a complex stellar kinematic
distribution function (Figure~2). In any case, this illustrates that
an adequate understanding of the very outer parts of the dSph is still
lacking, so that we are unable as yet to determine outer mass
distributions, or `total' masses. The situation in the inner regions
is however better defined and understood.

\begin{figure}[!h]
\begin{center}
\includegraphics[width=0.42\textwidth]{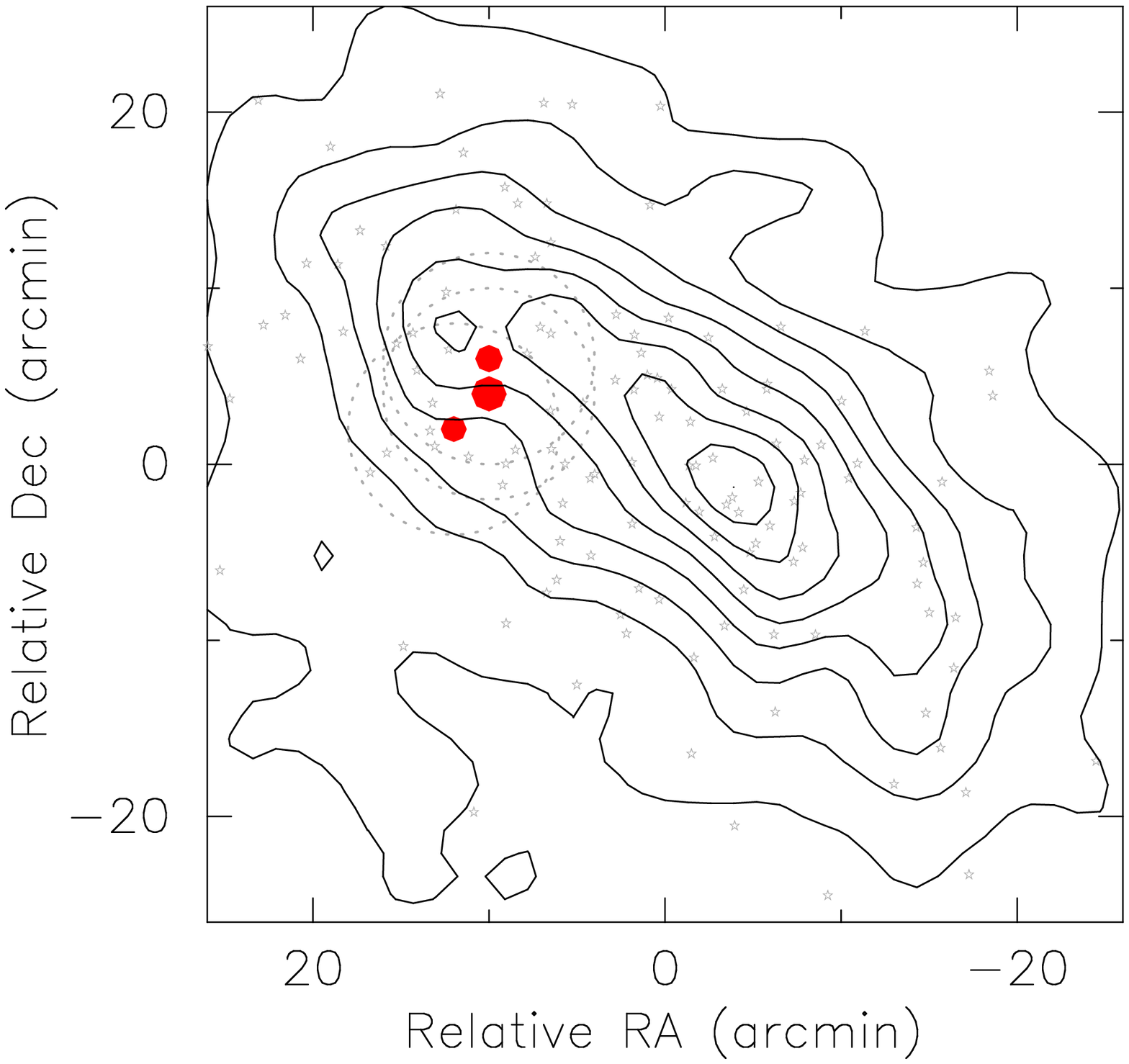}
\includegraphics[width=0.42\textwidth]{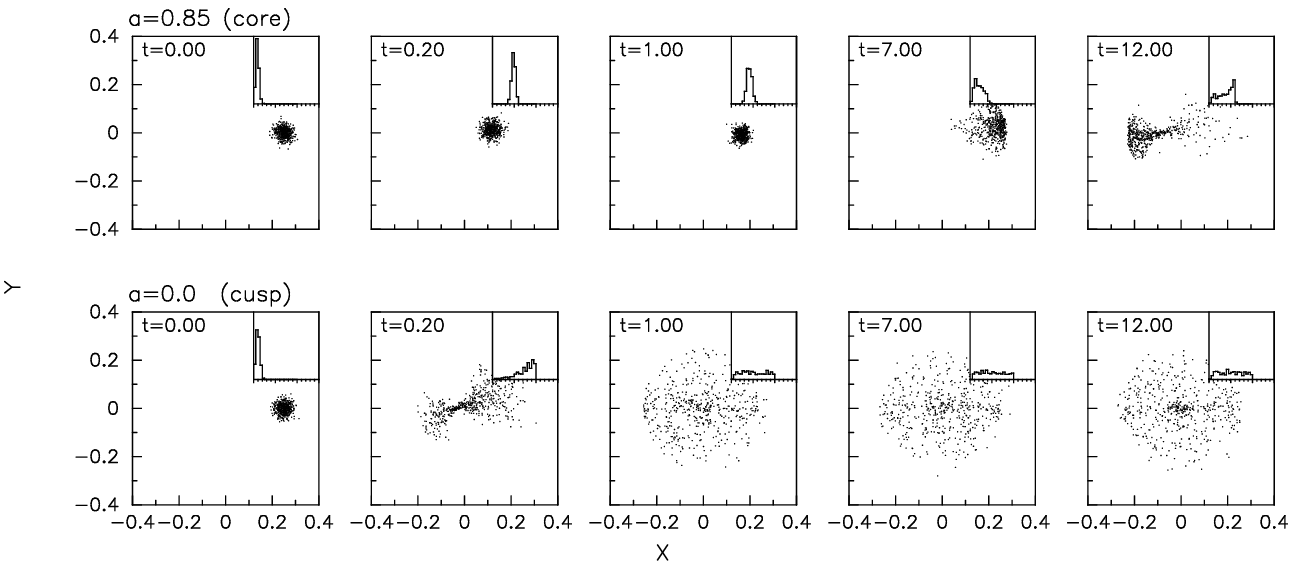}
\caption{TOP: Result of search for kinematic sub-populations in
UMi. Coutours are linearly spaced stellar isopleths; the second peak
of UMi's stellar population is visible at ${\rm RA}=12^\prime, {\rm
Dec}=8^\prime$ relative to UMi's center.  Gray stars are UMi RGB
member stars with measured velocities. The dots represent points where
a model with a kinematically cold sub-population is at least 1000
times more likely than a model composed of a single $8.8\,\rm
km\,s^{-1}$ Gaussian. BOTTOM: Simulation of
an unbound clump in a dark matter halo.  The halo has a density law
$\rho(r)\propto(a^2+r^2)^{-1/2}$; the time $t$ is in units of Gyr. Top panel: a
clump in a cored halo with $a=0.85$ persists for a Hubble time because
the potential is nearly harmonic.  Bottom panel: a clump in a cusped
potential ($a=0$) disrupts in less than 1 Gyr. The histograms at the
upper right corner of each snapshot show the distribution of total
velocity $v=(v_x^2+v_y^2+v_z^2)^{1/2}$; tick marks are spaced 1 $\rm
km\,s^{-2}$ apart.  The stars in a cored halo remain
coherent in velocity as well as in position. From
\cite{K03}.}
\end{center}
\end{figure}

Figure 3 summarises Jeans equation models for several of the dSph,
with in each case the simplest possible assumptions (isotropic
radially-constant velocity distribution). It is apparent that the
models are invalid at large radii, where an unphysical oscillation in
the mass profiles is evident. In the inner regions however the fit to
the data is good. In each case, a core-like mass distribution is
preferred. As noted above however, it is possible, by adding a
radially-variable stellar anisotropy (essentially an extra function)
to the fit, to fit steeper cusp-like central mass distributions.

Can one distinquish between shallow and steep density profiles using
other information?  Fortunately, in one special case, that of UMi, one
can. In UMi, an otherwise very simple system from an astrophysical
perspective, an extremely low velocity dispersion sub-structure exists
(figure 4a). We explain this \cite{K03} as a star cluster,
which has become gravitationally unbound (the normal eventual fate of
every star cluster, as mass is lost through normal stellar evolution),
and which now survives as a memory in phase space. Why does it survive
in configuration space? The group of stars have the same mean velocity
as the systemic velocity of UMi, so they must orbit close to the plane
of the sky, and hence through the central regions of UMi. As figure
(4b) illustrates, this is possible only if the tidal forces from the
UMi central mass gradient are weak. In fact, survival of this
phase-space structure in configuration space requires that UMi has a
cored mass profile. Similar results are provided by the survival of
the globular cluster system in the Fornax dSph \cite{G06}.

\subsection{Distribution Function models}

Standard cosmological simulations tend to overpredict the numbers of
low-mass haloes surrounding Milky Way-type galaxies compared to
observations. One possible solution (eg \cite{S02}) is that
all dwarf galaxies have very extended haloes, and are therefore very
much more massive than previously thought.  According to the Stoehr
etal analysis, the observed dispersion profile of Draco and Fornax can
be obtained by embedding the stellar distributions in $\Lambda$CDM
haloes of masses $\gtrsim 10^9$M$_\odot$, extending to several
kpc. While our existing data on Draco are compatible with this
hypothesis, it is not known whether it holds in general. 
More reliable masses and mass profiles are therefore desirable.  

The principal objective of our dynamical modelling work to date has
been to move beyond the traditional King models for dSphs, in order to
break the principal degeneracy of the problem, namely that a large
velocity dispersion at large radii in a dSph can be explained either
by the presence of large amounts of unseen mass at large radii, or by
tangential anisotropy of the velocity distribution. We have therefore
developed a family of dynamical models (described in detail in
\cite{W02}) which span the full range of possible halo models from
mass-follows-light to extended haloes. The models also incorporate
anisotropy in the velocity distribution. Our current modelling has
focussed on spherically symmetric dark haloes and stellar
distributions. Our analysis of our Draco data, together with extensive
Monte Carlo simulations, has demonstrated that data sets of several
hundred radial velocities are sufficient to discriminate between halo
models and to break the degeneracy between mass and velocity
anisotropy, provided these data are at sufficiently large radii
\cite{K01};\cite{W02}. The outermost data also place constraints on
any systemic rotation or tumbling which might contribute, along with
anisotropic kinematics, to the flattening seen in some dsph galaxies,
and provide further direct tests that a system is unaffected by tides,
so that an equilibrium dynamical analysis is appropriate.

The results of spherical modelling supports the basic results of the
Jeans' equation analysis, though of course use more information, so are
more robust. We are currently implementing new families of models to
allow more robust distinction between cored and cusped mass
distributions in the dSph.  One new option allowed by extant DF models
is to test the viability of the alternative gravity theory,
MOND. Wilkinson et al. have carried out such an analysis, concluding
that even with MOND dark matter is required. Available data for the
Fornax dSph can in fact be modelled quite well by MOND, assuming the
MOND scale parameter to be $a_0 = 2.1\times10^{-8} {\rm cm
s}^{-2}$. For UMi and Draco, however, even MOND requires either a high
M/L value ($\sim 20$), or a very different length scale parameter
($a_0 \sim 50\times10^{-8} {\rm cm s}^{-2}$).

\section{General results: systematic properties of dark matter on
small scales}

\begin{figure}[!h]
\begin{center}
\includegraphics[scale=0.4]{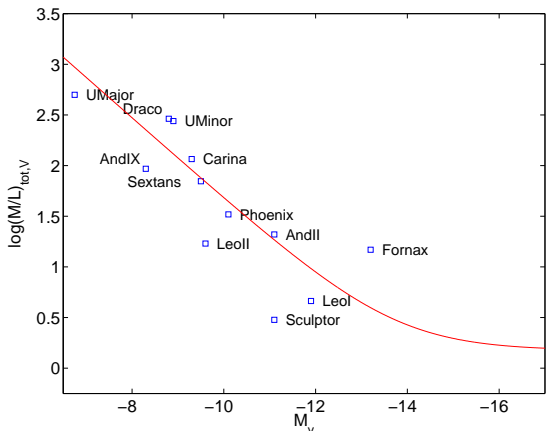}
\caption{Mass to light ratios {\it vs} galaxy absolute  V magnitude
for some Local Group dSph galaxies. The solid curve shows the relation
expected if all the dSph galaxies contain about $4\times10^7$ solar
masses of dark matter interior to their stellar distributions.}
\end{center}
\end{figure}

The profiles in figure 3, derived by Jeans' equation analyses, 
 illustrate two of our basic results. 
In every case, the simplest analysis favours cored mass
distributions. While cusped mass distribution can usually be fit to
the data, in at least one case, UMi, there is very strong direct
evidence that a cusp model is inadequate to explain all the available
information. The conservative assumption is therefore that all the
mass profiles are indeed cored, and are significantly shallower than
$r^{-1}$.

Secondly, all the dSph we have analysed to date show very similar, and
perhaps surprisingly low, central dark matter mass densities, with a
maximum value of $\sim 5 \times 10^8 M_{\odot} kpc^{-3}$, equivalent
to $\sim 20$GeV/cc. Interestingly, the rank ordering of the central
densities is in inverse proportion to system total luminosity, with
the least luminous galaxies being the most dense. This is of the
opposite sign to some CDM predictions, though we note a yet further
complication involving system mass below, since most CDM simulations
predict mass rather than luminosity.

We have not yet detected a reliable Keplerian decline, or any evidence
for tidal truncation, in any dSph galaxy dark halo studied so far. We
therefore (still) have no total mass determinations. Nonetheless, the
total masses determined out to the limits of the stellar distribution
show an interesting systematic effect. This is shown in Figure~5, a
plot whose style is adopted from \cite{M98}. Figure~5 displays
the relationship between the luminosity of the lowest-luminosity Local
Group galaxies against the derived (logarithmic) system mass-to-light
ratio. In this parameter space, the solid curve shows the relationship
anticipated if every dSph galaxy contains the same mass of dark matter
- in this case some $\sim4-5.10^7 M_{\odot}$ within the volume which
contains its stellar population. The system total luminosity is
measured directly.

It is apparent from Figure~5 that there is remarkably little spread in
mass apparent among the galaxies with absolute magnitute fainter than
$\sim -11.$ This relation was considered until recently to be a minor
curiosity, since it covered the dynamic range only from M$_V \sim -13$
to $-9$, a mere factor of forty or so in luminosity, and included only
8 galaxies. However, the recent analysis \cite{K05} of the
newly-discovered extremely low luminosity dSph galaxy UMa has extended
the validity of the relation by another two magnitudes, now a factor
of $\sim200$ in luminosity, and to total mass-to-light ratios in
excess of 1000. Several other new very low luminosity dSph satellite
galaxy candidates have been discovered in the last few months, while
new studies of several known galaxies have recently been
completed \cite{K06}. It will be very interesting to see if these dynamical
studies strengthen or disprove this apparent trend.


\begin{thebibliography}{18}

\bibitem{BT}
Binney, J \& Tremaine, S 1987 Galactic Dynamics (Princeton: PUP)
\bibitem{G06}
Goerdt etal 2006 MNRAS 368 1073
\bibitem{Gr}
Grillmair etal 1998 AJ 115, 144
\bibitem{K01}
Kleyna et al. 2001 ApJ 564 L115
\bibitem{K03}
Kleyna etal 2003 ApJ 588 L21
\bibitem{K05}
Kleyna, J., Wilkinson, M, Evans, N. Wyn, Gilmore, G
2005 ApJ 630 L141	
\bibitem{K06}
Koch, A etal AJ in press 2006
\bibitem{KG89}
Kuijken, K., \& Gilmore, G. 1989 MNRAS 239 605
\bibitem{KG91}
Kuijken, K., \& Gilmore, G. 1991 ApJ 367 L9
\bibitem{M98}
Mateo etal 1998 AJ 116 2315
\bibitem{S02}
Stoehr et al. 2002 MNRAS 335 L84
\bibitem{rg}
Read, J and Gilmore, G.  2005 MNRAS 356 107
\bibitem{R06}
Read, J. I.; Wilkinson, M. I.; Evans, N. Wyn; Gilmore, G.;
Kleyna, Jan T.  2006 MNRAS 367 387	
\bibitem{T02}
Tremaine etal 2002 ApJ 574, 740
\bibitem{W02}
Wilkinson et al. 2002 MNRAS 330 778
\bibitem{W04}
Wilkinson etal 2004 ApJ 611 L21
\bibitem{Wy}
Wyse etal, NewAstr 7 395 2002
\end{thebibliography}
\end{document}